\documentclass[aps,prb,twocolumn,superscriptaddress,longbibliography]{revtex4-1}

\usepackage{graphicx}
\usepackage{array}
\usepackage{dcolumn}
\usepackage{bm}
\usepackage{float}
\usepackage{enumitem}
\usepackage[colorinlistoftodos, textwidth=4cm, shadow]{todonotes}
\newcommand{\degrees}{$^{\circ}$}

\begin{document}

\thispagestyle{empty} 
Notice: This manuscript has been authored by UT-Battelle, LLC under Contract No. DE-AC05-00OR22725 with the U.S. Department of Energy. The United States Government retains and the publisher, by accepting the article for publication, acknowledges that the United States Government retains a non-exclusive, paid-up, irrevocable, world-wide license to publish or reproduce the published form of this manuscript, or allow others to do so, for United States Government purposes. The Department of Energy will provide public access to these results of federally sponsored research in accordance with the DOE Public Access Plan (http://energy.gov/downloads/doe-public-access-plan).

\title{Magnetism of Nd$_2$O$_3$ single crystals near the N\'{e}el temperature}

\author{Binod K. Rai}
\email[]{rairk@ornl.gov}
\affiliation{Materials Science and Technology Division, Oak Ridge National Laboratory, Oak Ridge, TN 37831, USA}

\author{A. D. Christianson}
\affiliation{Materials Science and Technology Division, Oak Ridge National Laboratory, Oak Ridge, TN 37831, USA}

\author{G. Sala}
\affiliation{Neutron Scattering Division, Oak Ridge National Laboratory, Oak Ridge, TN 37831, USA}

\author{M. B. Stone}
\affiliation{Neutron Scattering Division, Oak Ridge National Laboratory, Oak Ridge, TN 37831, USA}


\author{Y. Liu}
\affiliation{Neutron Scattering Division, Oak Ridge National Laboratory, Oak Ridge, TN 37831, USA}

\author{A. F. May}
\email[]{mayaf@ornl.gov}
\affiliation{Materials Science and Technology Division, Oak Ridge National Laboratory, Oak Ridge, TN 37831, USA}

\date{\today}

\begin{abstract}
Single crystals of Nd$_2$O$_3$ were grown and characterized using neutron scattering and thermodynamic measurements.  Nd$_2$O$_3$ has long-range antiferromagnetic order below $T_{\rm N}$ = 0.55\,K and specific heat measurements have demonstrated that a significant amount of the magnetic entropy is released above $T_{\rm N}$.  Inelastic neutron scattering experiments reveal a magnetic mode(s) with little dispersion peaked at $\approx$\,0.37\,meV that is of greatest intensity below $T_{\rm N}$ but persists above 2$T_{\rm N}$.  This persistence of dynamic correlations is likely related to frustrated interactions associated with the nearly-ideal stacked triangular lattice geometry of $J_{\textrm{eff}}$\,=\,1/2 spins on Nd$^{3+}$ ions.  The magnetization is observed to be strongly anisotropic at all temperatures due to crystal field effects, with easy-plane anisotropy observed.  A non-compensated magnetic structure is inferred from the temperature-dependence of the magnetization when a magnetic field of sufficient strength is applied within the basal plane near $T_{\rm N}$, and the evolution  of the long-range order is summarized in a temperature-field phase diagram.

\end{abstract}

\maketitle

\section{Introduction}
Magnetic interactions in stacked triangular lattice systems can be highly frustrated, with the resulting ground state dependent on the nature of the interactions and stacking sequence\cite{Liu2016}. In particular, the hexagonal close packed (HCP) lattice (ABAB stacking) of triangular nets can lead to a variety of magnetic states, including a classical spin liquid\cite{Hoang2012,Liu2016}. Identifying relevant materials with this type of three-dimensional frustration and probing the impacts of frustration on the magnetism motivates this work.

\begin{figure}[b!]
	\includegraphics[clip,width=\columnwidth]{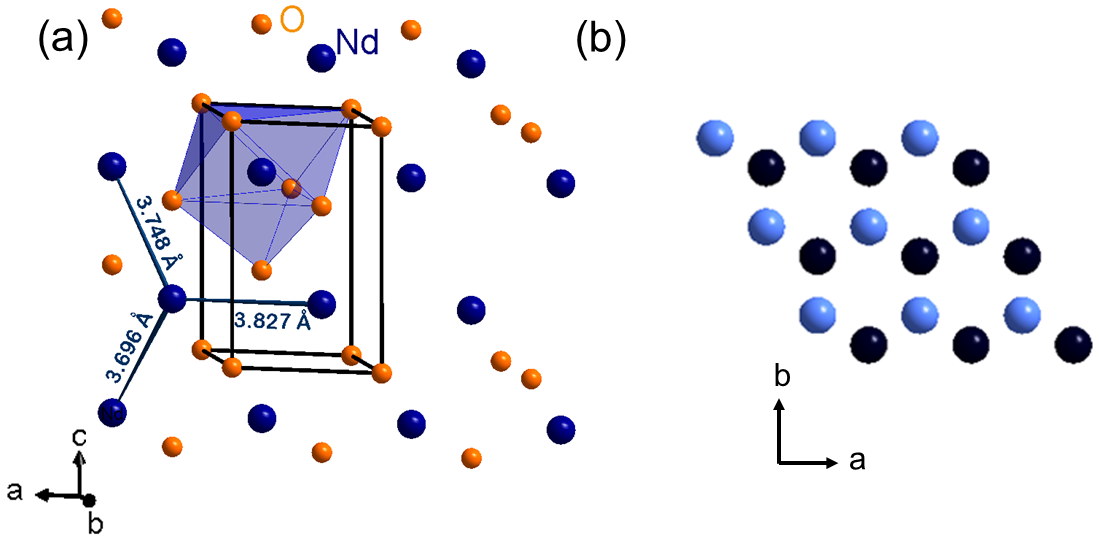}
\caption{\label{Structure} (a) Crystal structure of Nd$_2$O$_3$ with space group $P\bar{3}m1$. The shortest Nd-Nd distances are indicated, one unit cell is outlined, and the coordination polyhedra with seven oxygen ions around Nd$^{+3}$ is shown. (b) View down [001] illustrating the triangular arrays of the Nd atoms, with different colors indicating the different layers (different z-coordinates).}
\end{figure}

Nd$_2$O$_3$ is an insulating material that contains ABAB stacking of triangular nets of Nd$^{3+}$ ions and a $J_{eff}$\,=\,1/2 ground state\cite{Pauling1928,Boucherle1975, Boldish1979,Faucher1982,Gruber2002,Sala2018}. The crystal structure of Nd$_2$O$_3$ is shown in Fig.\,\ref{Structure}, with pertinent Nd-Nd distances provided; this is a trigonal material with centrosymmetric space group $P\bar{3}m1$.  Seven oxygen atoms coordinate the Nd ions and the three nearest neighbor Nd-Nd distances are observed to be similar. The longest of these distances is within the basal plane, where the Nd$^{3+}$ ions form triangular nets.  These triangular nets are stacked in a close-packed manner along the $c$-axis, and the shortest Nd-Nd distances involve traversing out of the basal plane.  The resulting lattice of Nd ions is similar to an HCP lattice with minor distortions.

The magnetic properties of Nd$_2$O$_3$ were recently investigated using polycrystalline samples.\cite{Sala2018}  Nd$_2$O$_3$ was found to possess long-range antiferromagnetic (AFM) order below a N{\'e}el temperature of $T_\mathrm{N}$\,=\,0.55\,K\cite{Sala2018}. Neutron powder diffraction data revealed a magnetic propagation vector of $\vec{k}\,=\,(\frac{1}{2},0,\frac{1}{2})$, which indicates anti-parallel alignment of moments along the $a$-axis and along the $c$-axis\cite{Sala2018}.  Two spin structures were found to be consistent with the data, and they differ only in the sequence for the anti-parallel alignment of the staggered Nd ions when moving along the $c$-axis with either ($--++$) or ($+--+$) stacking.  In these spin structures, the ordered moments (1.87(10) $\mu_B$/Nd at $T$\,=\,0.28\,K) are lying within the $ab$-plane.  Inelastic neutron scattering data suggest the anisotropy originates from crystal electric field (CEF) effects.\cite{Sala2018}  There are four crystal field excitations from the ground state, spanning the energy range of 2.8--60\,meV ($\approx$\,32-700\,K)\cite{Henderson1967,Caro1979,Sala2018}.  Nd$_2$O$_3$ thus possesses a Kramer's doublet ground state characterized by an effective spin state $J_{eff}$\,=\,$\frac{1}{2}$ that is isolated from excited states for temperatures below $\approx$\,5\,K.  A broad Schottky anomaly is observed near 12\,K and associated with the first excited crystal field state.

The specific heat of Nd$_2$O$_3$ has a sharp $\lambda$ anomaly near $T_{\rm N}$\,=\,0.55\,K and a broad peak near 1.2\,K\cite{Sala2018}.  The $\lambda$ anomaly only accounts for about 30\% of the expected $R$ln(2) entropy of the ground state, implying that short-range magnetic correlations persist to well above $T_{\rm N}$.  Such behavior may evidence that Nd$_2$O$_3$ hosts frustrated interactions associated with its stacked triangular lattices.  The importance of exchange frustration and the role of static and/or dynamic spin-spin correlations is best assessed by studying single crystal specimens, which allow magnetic anisotropy to be characterized and can enhance the utility of spectroscopic tools such as inelastic neutron scattering.

This study reports on the growth of Nd$_2$O$_3$ single crystals and their characterization through magnetization, specific heat and neutron scattering measurements.  An anisotropic response of the magnetization $M$ to an applied field $H$ is observed above and below $T_{\rm N}$.  Temperature-dependent magnetization data down to $T$\,=\,0.4\,K suggest that a non-compensated magnetic structure (canted AFM) exists for $H$\,$\perp$\,$c$ when $\mu_0H$\,$>$\,0.4\,T. On the contrary, the phase transition associated with the long-range AFM order is continually suppressed with increasing field for $H$\,$\parallel$\,$c$.  Neutron scattering experiments were undertaken to investigate the nature of the short-range correlations that persist above $T_{\rm N}$.  Diffraction-based evidence for static short-range correlations above $T_{\rm N}$ was not found.  However, a correlated low-energy magnetic excitation with very little dispersion was found to persist above $T_{\rm N}$.  These data reveal that Nd$_2$O$_3$ possesses geometric frustration and/or competing ground states that result in correlated dynamics above $T_{\rm N}$, and thus a detailed investigation of the nature of the spin-spin interactions is warranted.

\section{Experimental Methods}

\begin{figure}[b!]
	\includegraphics[width=0.9\columnwidth]{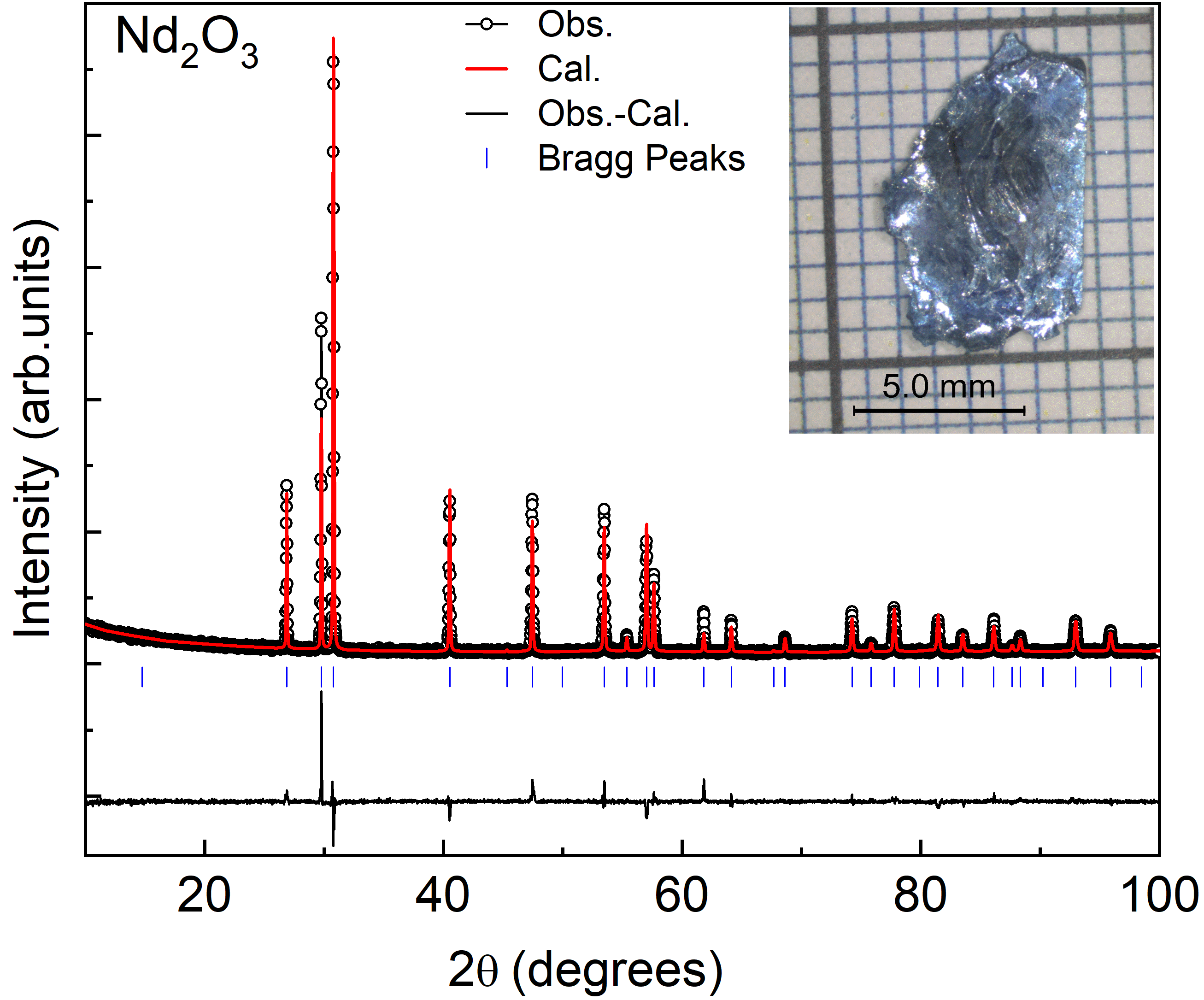}
\caption{\label{XRD} Room temperature x-ray powder diffraction data for a pulverized crystal (symbols) together with the Rietveld refinement (red line) and Bragg positions (vertical lines). A picture of a cleaved crystal is shown in the inset.}
\end{figure}

Single crystals of Nd$_2$O$_3$ were grown in an optical floating-zone furnace under flowing oxygen. Dried Nd$_2$O$_3$ powders (99.99$\%$) were formed into rods that were sintered in air at $T$\,=\,1100\,$^{\circ}$C for two days. The growths utilized a four-mirror, xenon-lamp furnace (Crystal System Corporation, Model FZ-T-12000-X-VPM-PC-OR). The growths resulted in crystals that were easily cleaved to plates with $c$-axis normal. A representative crystal after cleaving is shown in the inset of Fig. \ref{XRD}. We note that the growth was not particularly stable, seemingly due to a low viscosity.  Single crystals of Nd$_2$O$_3$ have been previously grown using the Verneuil technique,\cite{Bernier1973} in which the melt drops onto a crystalline boule.  Nd$_2$O$_3$ crystals or powders exposed to air eventually turn into neodymium hydroxide powders, and thus we minimized crystal exposure time in air by quickly loading samples for measurements and storing samples under dynamic vacuum with desiccant before and after measurements. 

X-ray diffraction data were collected in a PANalytical X'Pert Pro MPD diffractometer (Cu K$\alpha_1$ radiation). The x-ray diffraction data were analyzed with the program FullProf\cite{FullProf} using the Rietveld technique. Temperature-dependent magnetization measurements down to 2\,K were performed in a field-cooled condition using a Quantum Design (QD) Magnetic Property Measurement System. The $M$($T$) data from 0.4\,K to 1.8\,K were obtained upon warming in a field-cooled condition using the iHe3 insert in a QD MPMS3 and the sample was fixed to the measurement holder using Dow Corning high-vacuum grease.  Isothermal magnetization data were collected upon increasing the magnetic field. Specific heat data down to $T$\,=\,0.4\,K were obtained using a $^3$He insert in a QD Physical Property Measurement System. Specific heat measurements with the field applied in the $ab$-plane were collected using a sapphire plate to aid sample alignment, with appropriate addenda collected.

Elastic neutron scattering measurements were performed using the CORELLI diffuse scattering spectrometer\cite{Feng2018} at the Spallation Neutron Source (SNS) at Oak Ridge National Laboratory (ORNL). A Nd$_2$O$_3$ single crystal of mass $\approx$\,2\,g was mounted on a copper plate and wrapped in copper foil to enhance thermal transport. This assembly was attached to a copper pin at the bottom of a helium-3 cryostat, which provided a base temperature of 0.25\,K. The sample was mounted with the ($h$ 0 $l$) plane horizontal, and the experiments were conducted by rotating the crystal through $\approx$\,200\degrees ~in 2\degrees~steps at multiple temperatures below and above $T_{\rm N}$.  The inelastic spectra were measured at CNCS using a dilution refrigerator insert in a 8\,T superconducting magnet.  An incident energy of 12\,meV was utilized to align the sample and obtain the UB matrix, and data collections occurred at 1.72\,meV to study the low lying excitations.  Data were collected using a single crystal aligned in the ($h$ 0 $l$) scattering plane.  The crystal was measured over a 180\degrees rotation about the vertical axis.

\section{Results and Discussion}

\subsection{Crystallography from Diffraction}

The phase purity and crystal quality of the Nd$_2$O$_3$ single crystals were verified using powder x-ray diffraction and neutron single crystal elastic scattering.  Room temperature x-ray powder diffraction data from pulverized single crystals of Nd$_2$O$_3$ are shown in Fig. \ref{XRD}. The Rietveld refinement of the powder diffraction data are well-modeled by the reported crystal structure with trigonal space group $P\bar{3}m1$ (No. 164). The lattice parameters obtained from refinement are $a$\,=\,3.8299(1)\,\AA\, and $c$\,=\,5.9995(2)\,\AA\,, which are in agreement with our previous neutron powder diffraction data\cite{Sala2018}.

\begin{figure}[t!]
	\includegraphics[width=0.9\columnwidth]{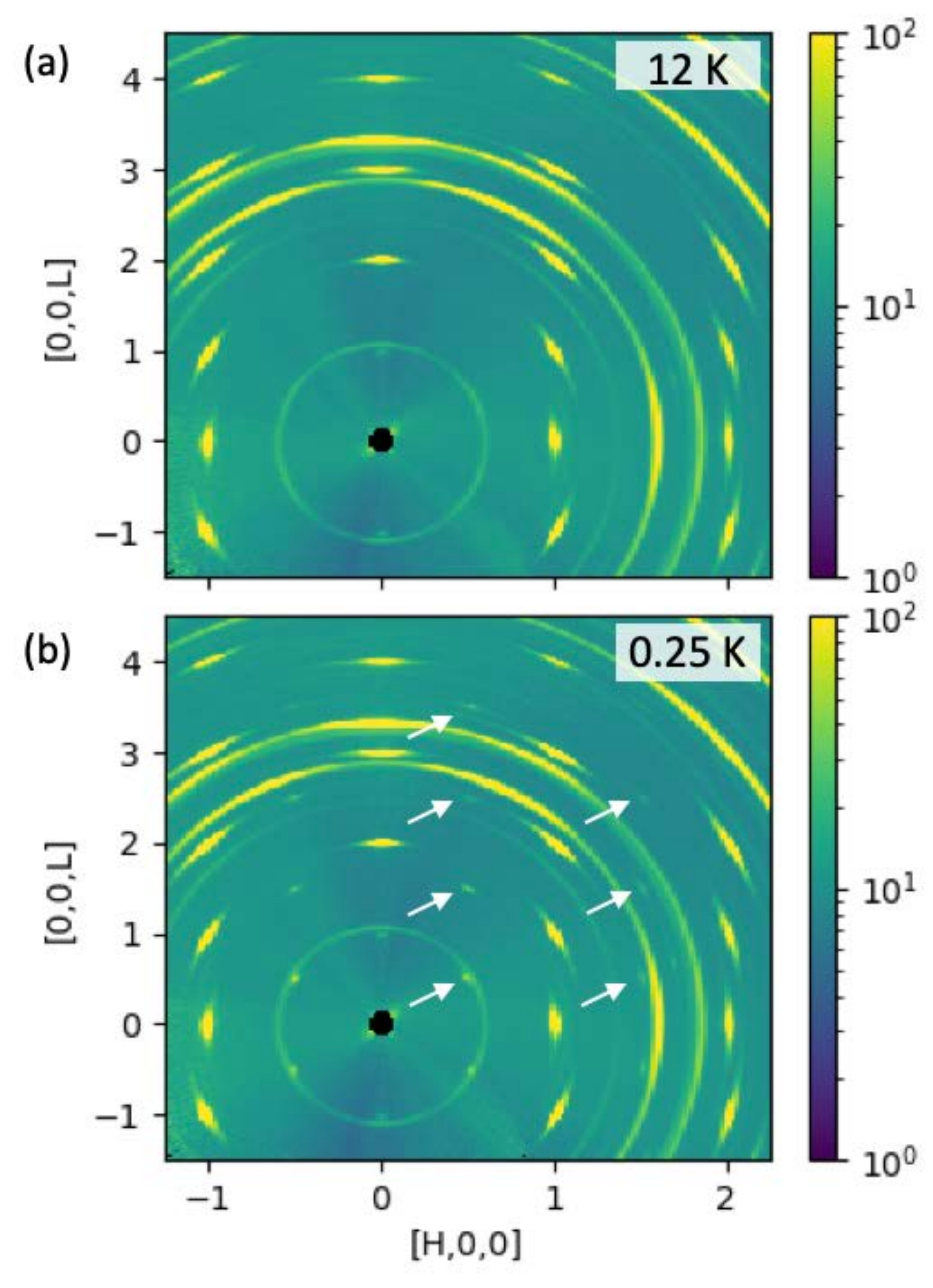}
\caption{\label{CORELLI} Neutron diffraction patterns in ($h 0 l$) scattering plane obtained using a single crystal on the CORELLI instrument at the SNS at (a) $T$\,=\,12\,K and (b) 0.25\,K.  The white arrows in (b) indicate the location of magnetic scattering from long-range antiferromagnetic order in Nd$_2$O$_3$.}
\end{figure}

Neutron single crystal diffraction experiments were performed at the CORELLI beamline at the SNS.  Data were collected at 0.25\,K$\leq$\,$T$\,$\leq$12\,K with the sample aligned in the ($h$ 0 $l$) scattering plane.  The observed Bragg peaks were sharp, demonstrating that the crystal is of high quality.  Figure \ref{CORELLI}(a) displays the diffraction pattern for $T$\,=\,12\,K (in the paramagnetic phase). As shown in Fig.\,\ref{CORELLI}(b), the diffraction data at $T$\,=\,0.25\,K contain extra reflections due to the antiferromagnetic order and these are indexed using a propagation vector of $\vec{k}\,=\,(\frac{1}{2},0,\frac{1}{2})$ as previously reported.\cite{Sala2018} The location of this magnetic scattering is indicated by the white arrows. Magnetic diffraction peaks were not observed at or above $T$\,=\,0.65\,K. The ring of diffraction intensity closest to $Q$\,=\,0 was indexed as coming from Nd(OH)$_3$, likely due to degradation of the sample surface during loading; additional powder rings also originate from the sample can.

\subsection{Anisotropy and Magnetic Phase Diagram}

\begin{figure}[t!]
\includegraphics[width=0.9\columnwidth]{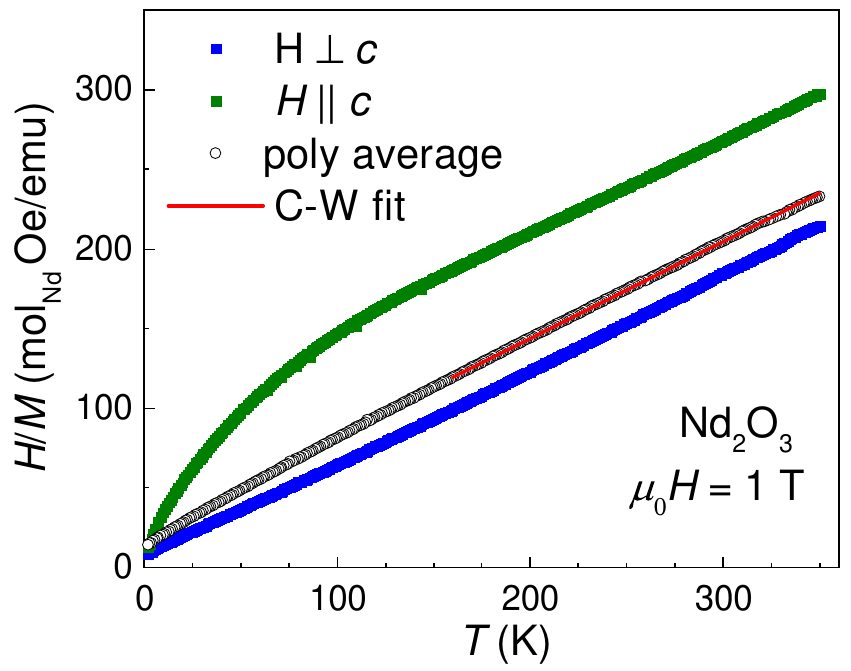}
\caption{\label{Curie} The inverse magnetic susceptibility ($1/\chi\,=\,H/M$) of Nd$_2$O$_3$ for different orientations of the applied field as well as the polycrystalline average, as indicated in the legend.  The Curie Weiss model (`C-W fit') was obtained by fitting data between 160 and 350\,K.}
\end{figure}

Magnetization ($M$) measurements revealed a strong magnetic anisotropy in Nd$_2$O$_3$.  The inverse of the magnetic susceptibility ($1/\chi\,=\,H/M$) is plotted in Fig.\,\ref{Curie}, where a large difference is observed based on the orientation of the applied field $H$. These data are consistent with the literature and reveal easy-plane anisotropy (moments perpendicular to [001]). This anisotropy is driven by the crystal electric fields (CEF)\cite{Tueta1972,Bernier1973,Caro1979,Sala2018}.  The CEF spectrum of Nd$_2$O$_3$ consists of doublets at 0, 2.86(3), 10.49(2), 30.46(3), and 60.29(4)\,meV\cite{Sala2018}.  The anisotropic $g$-factors estimated from the CEF model developed using neutron scattering data are $g_{xy}$\,=\,1.733 and $g_z$\,=\,0.213. Furthermore, the majority (67\%) of the Kramer's doublet ground state is composed of $J_z$\,=\,$\frac{1}{2}$\cite{Sala2018}.

The susceptibility follows Curie-Weiss behavior above $\approx$\,160\,K, as demonstrated by the linear temperature dependence of $H/M$ in Fig.\ref{Curie}. The data were fit to the Curie-Weiss model ($\chi$\,=\,$M/H$\,=\,$C$/($T-\theta$)), where the Curie constant $C$ is related to the effective moment and $\theta$ is the Weiss temperature.  The fit was conducted between 160 and 350\,K using the polycrystalline average $M_{ave}$\,=\,$(2M_{ab}+M_c)/3$. This yielded an effective moment of $\mu_{eff}$\,=\,3.63\,$\mu_B$/Nd, consistent with the 3.62\,$\mu_B$ expected for a Nd$^{3+}$ free ion.  The $\theta$ from the fit is -35\,K, suggesting the dominance of AFM correlations. However, this value is seemingly impacted by crystal field effects.  Single ion anisotropy can result in large effective Weiss temperatures and that is likely the case here.\cite{Johnson2017}  The presence of the first crystal electric field excited state near 2.86\,meV makes it unreasonable to fit the low-$T$ data to a Curie Weiss model.  For this reason, the strength of the exchange interactions can only be accurately assessed through probes that examine the spin-spin correlations directly, such as inelastic neutron scattering.

\begin{figure}[h!]
    \includegraphics[width=0.9\columnwidth]{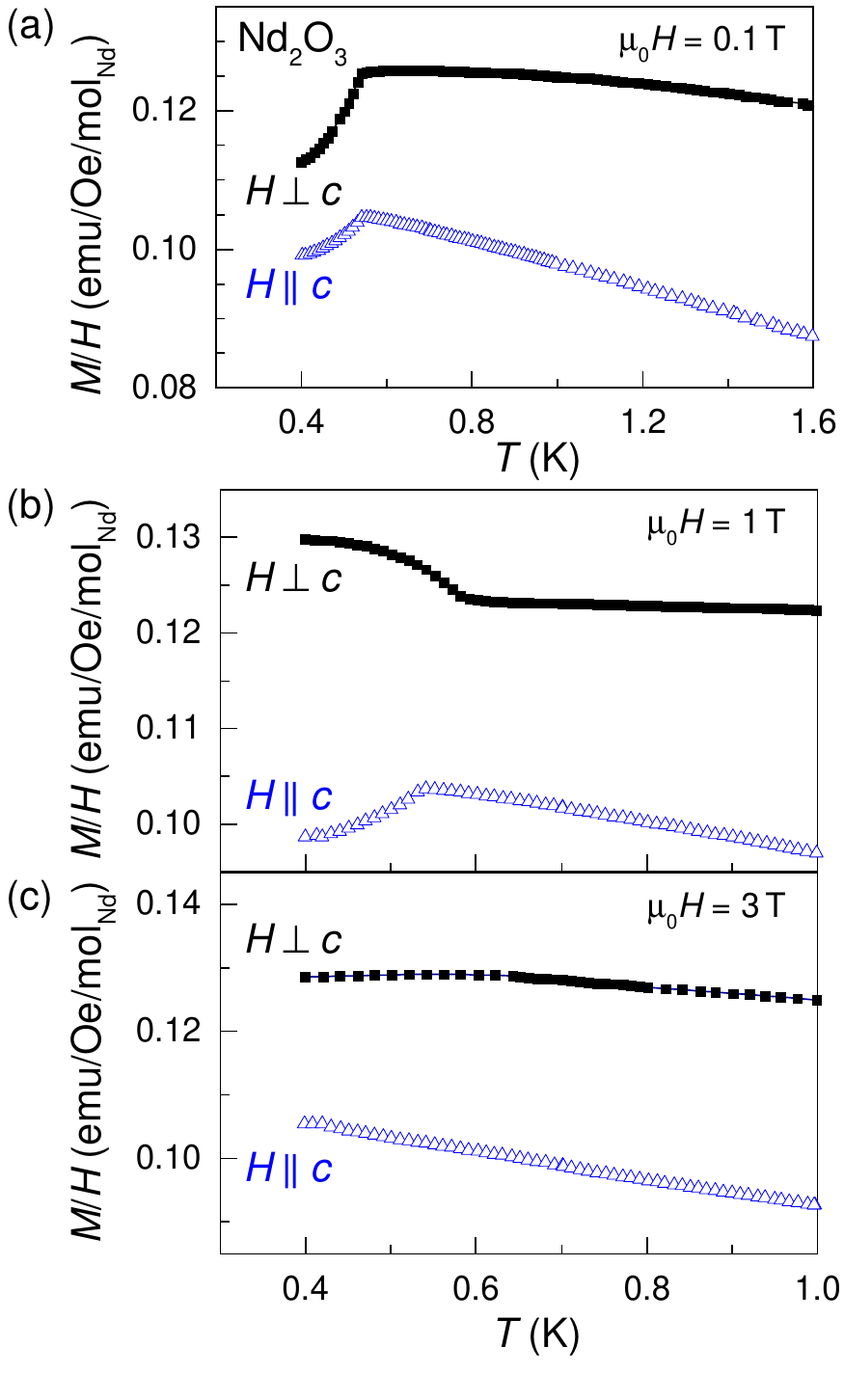}
\caption{\label{MT_1} The temperature-dependent magnetization data of Nd$_2$O$_3$ near $T_{\rm N}$\,=\,0.55\,K with different field orientations labeled for (a) $\mu_0H$\,=\,0.1\,T, (b) $\mu_0H$\,=\,1.0\,T and (c) $\mu_0H$\,=\,3.0\,T.}
\end{figure}

The magnetization of Nd$_2$O$_3$ single crystals was measured down to $T$\,=\,0.4\,K to characterize the magnetic anisotropy near $T_{\rm N}$.  As shown in Fig.\,\ref{MT_1}(a), $M$($T$) has a cusp near $T_{\rm N}$\,=\,0.55\,K for both orientations of the applied field for $\mu_0H$\,=\,0.1\,T; similar behavior was also observed for lower applied fields. This cusp corresponds to the onset of long-range AFM order upon cooling. The observed value of $T_\mathrm{N}$ is consistent with that obtained from neutron diffraction and specific heat measurements previously reported for polycrystalline samples\cite{Sala2018}. The magnetization is larger for $H$\,$\perp$\,$c$ than for $H$\,$\parallel$\,$c$ and the data for $H$\,$\perp$\,$c$ are fairly temperature independent in the region just above $T_{\rm N}$.

The magnetic response is qualitatively different for the two orientations when the applied field is greater than $\approx$\,0.3\,T.  This is illustrated in Fig.\,\ref{MT_1}(b) for an applied field of 1\,T, where it is observed that the in-plane magnetization increases sharply upon cooling starting slightly above $T_{\rm N}$.  By contrast, the data for $H$\,$\parallel$\,$c$ are as expected for an antiferromagnet, with a cusp feature near $T_{\rm N}$ being suppressed to lower $T$ with increasing applied field (see also Fig.\,\ref{MT_1}(c)).  Based on the $M$($T$) data, the critical field to suppress $T_\mathrm{N}$ beyond 0.4\,K for $H$\,$\parallel$\,$c$ is $\mu_{0}H\,\approx$\,2.5\,T.

\begin{figure}[h!]
    \includegraphics[width=0.92\columnwidth]{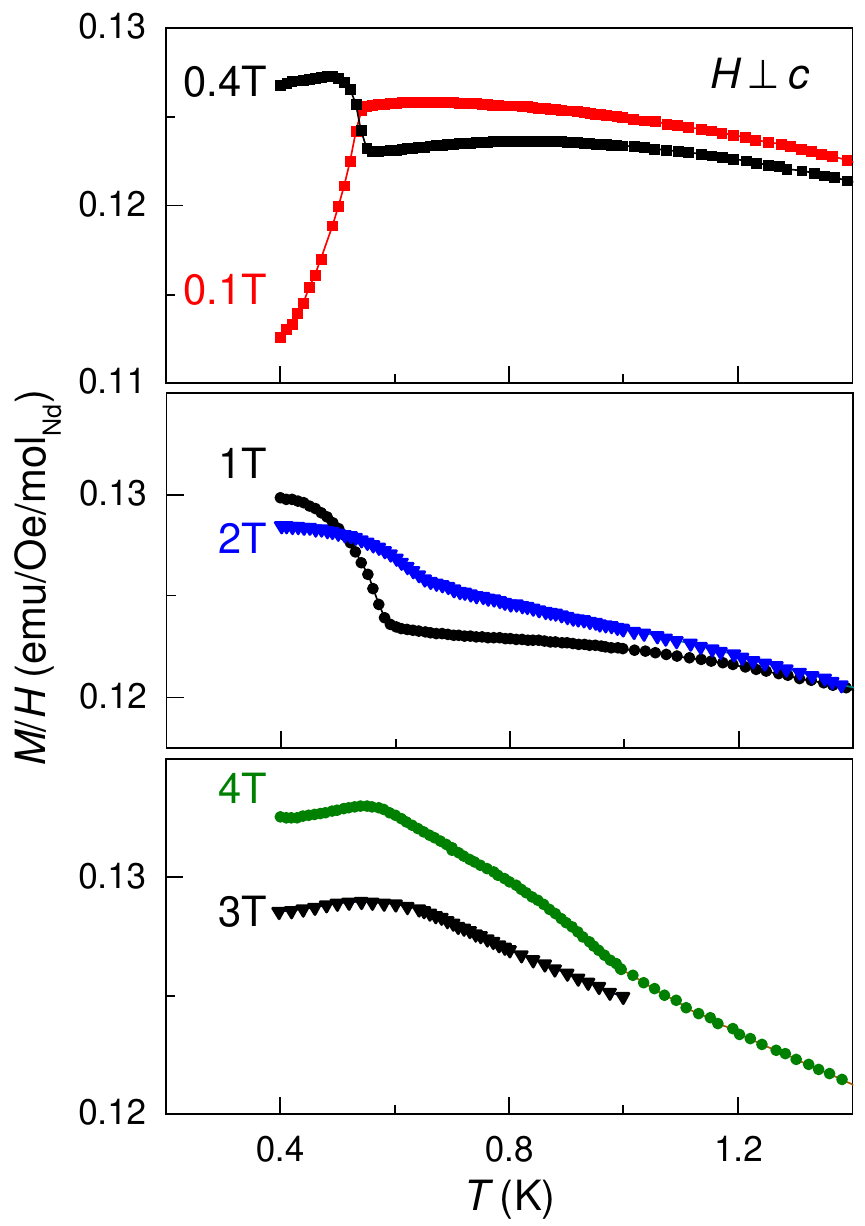}
\caption{\label{Hperp} Magnetization data for $H$\,$\perp$\,$c$ displaying the evolution of $M$($T$) with increasing applied field moving from the top to bottom panel.}
\end{figure}

The evolution of $M$($T$) for fields applied within the basal plane is illustrated in Fig.\,\ref{Hperp}. The upturn is rather sharp for an applied field of 0.4\,T.  The response for fields between 0.2-0.4\,T is characterized by a competition between these two behaviors, with a maximum in $M$($T$) observed just below the upturn.  For an applied field of 1\,T within the basal plane, the behavior of $M$($T$) is similar to that of a ferromagnet with a critical temperature near 0.6\,K.  The net increase in $M$ upon cooling is rather small, and thus a ferromagnetic description does not seem warranted.  Rather, we speculate that a canted antiferromagnetic spin structure with a small uncompensated moment is formed in response to these in-plane applied fields. We label this non-compensated AFM state `AFM$_C$' and the zero-field spin configuration `AFM$_0$'.  Upon further increase of the applied field, the onset of the ferromagnetic-like $M$($T$) shifts to higher temperature.  At still higher fields, the transition is smeared out in $M$($T$) and the system seemingly returns to a (polarized) paramagnetic state.  The critical field to obtain $M$($T$) behavior characteristic of the AFM$_C$ phase is slightly sample dependent (0.3-0.4\,T) but the overall $H$-$T$ response of the magnetization is similar in all samples examined.   The broad feature observed for 4\,T near 0.85\,K is perhaps an artifact associated with the iHe3 system's temperature control at the time of the measurement.

\begin{figure}[h!]
    \includegraphics[width=0.92\columnwidth]{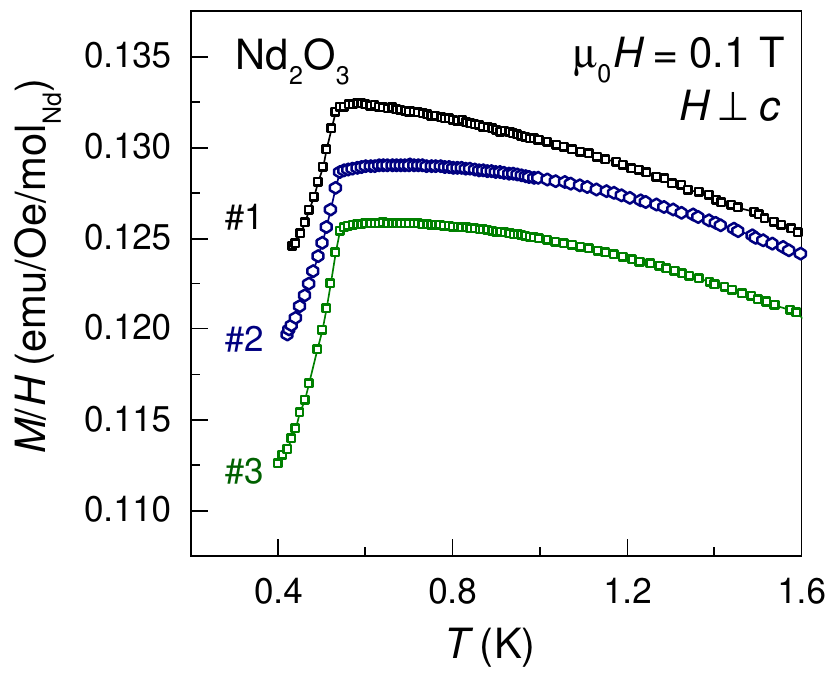}
\caption{\label{MT_2} The temperature-dependent magnetization of Nd$_2$O$_3$ for three separate samples from the same crystal at $\mu_0 H$\,=\,0.1\,T for $H$\,$\perp$\,$c$.}
\end{figure}

\begin{figure}[t!]
    \includegraphics[width=0.92\columnwidth]{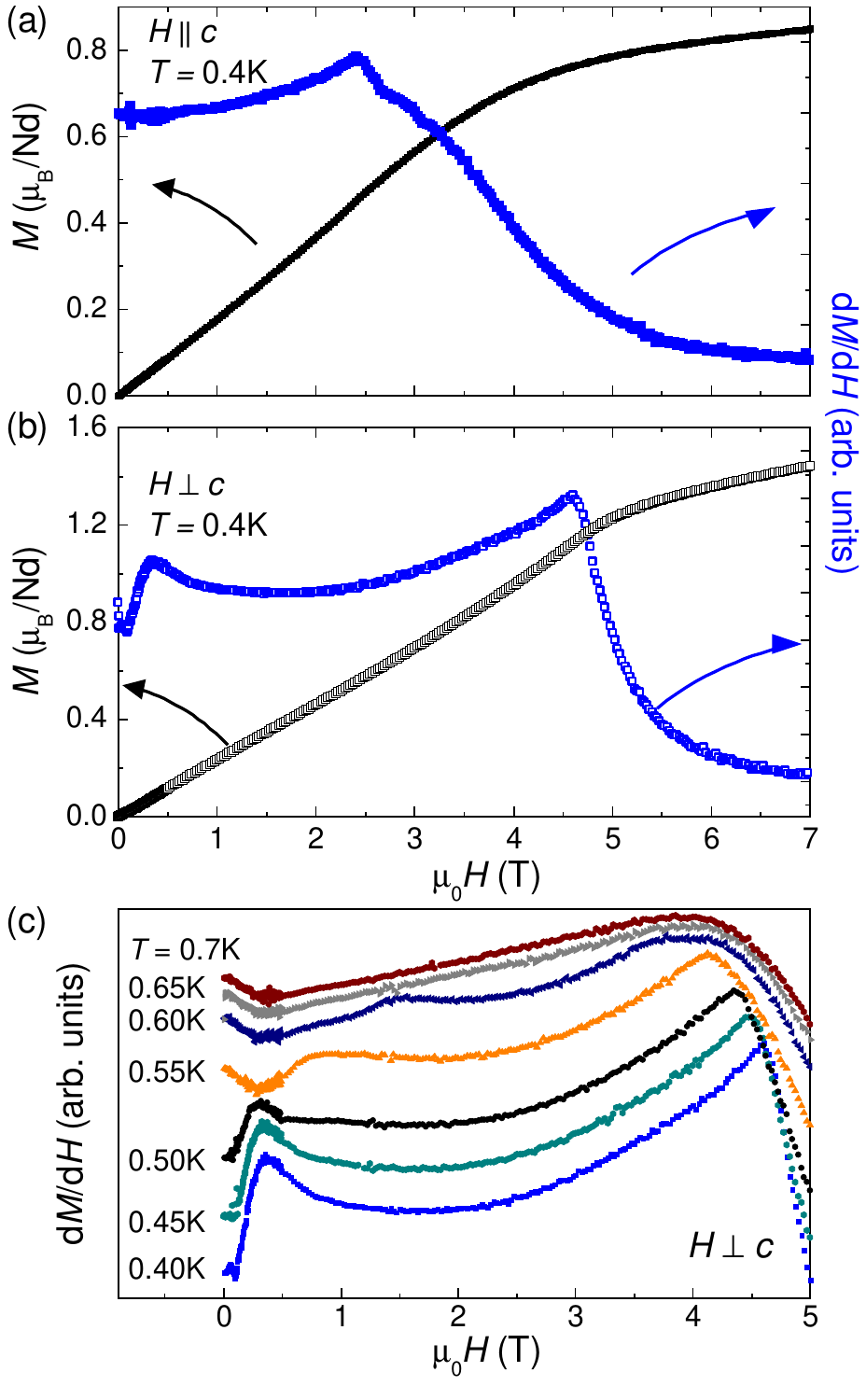}
\caption{\label{MH} Isothermal magnetization (left) and its derivative (right) measured below $T_\mathrm{N}$ for (a) $H$\,$\parallel$\,$c$ and (b) $H$\,$\perp$\,$c$. (c) d$M$/d$H$ for $H$\,$\perp$\,$c$ at various temperatures near $T_{\rm N}$ with data shifted vertically for clarity.}
\end{figure}

A broad maximum just above $T_\mathrm{N}$ is observed in $M$($T$) when $H$\,$\perp$\,$c$, which is qualitatively different from $M$($T$) for $H$\,$\parallel$\,$c$.  This disparity is observed in Fig.\ref{MT_1}(a). The broad maximum becomes more apparent for intermediate fields, as shown in Fig.\,\ref{Hperp} where a maximum is centered near 0.8\,K for 0.4\,T. Similar behavior was observed in different samples, and low-field data are shown in Fig.\,\ref{MT_2} for three pieces cleaved from the same crystal.  Due to the enhanced magnetization below $T_{\rm N}$ for intermediate fields (canted state), it is difficult to determine how the broad maximum evolves with increasing $H$.   These $M$($T$) data suggest that there might be short-range correlations with easy-plane character above $T_{\rm N}$; this anisotropy is consistent with that imposed by the crystal field effects.

The isothermal magnetization data $M$($H$) below $T_{\rm N}$ have a non-linear response to the applied field, as shown in Fig.\,\ref{MH}. The  magnetization is larger for $H$\,$\perp$\,$c$ than for $H$\,$\parallel$\,$c$, consistent with the temperature-dependent data and the anisotropic $g$-factors.\cite{Sala2018}  The induced moments are 0.85 and 1.43 $\mu_B$/Nd for $H$\,$\parallel$\,$c$ and $H$\,$\perp$\,$c$, respectively at $\mu_{0}H$\,=\,7\,T and $T$\,=\,0.4\,K.

The derivatives d$M$/d$H$ are shown on the right axis of Fig.\,\ref{MH}(a,b) for $T$\,=\,0.4\,K, as well as for several temperatures for $H$\,$\perp$\,$c$ in Fig.\,\ref{MH}(c).  As seen in Fig.\,\ref{MH}(a), d$M$/d$H$ has a cusp around $\mu_{0}H\,\approx$\,2.4\,T for $H$\,$\parallel$\,$c$ at $T$\,=\,0.4\,K. Considering also the $M$($T$) data (Fig.\,\ref{MT_1}), this corresponds to the field required to suppress long-range AFM order to below 0.4\,K for $H$\,$\parallel$\,$c$, i.e. the critical field to drive between the AFM$_0$ ground state and a (partially-polarized) paramagnetic phase at 0.4\,K.  The likely existence of short-range correlations above $T_\mathrm{N}$ complicates this description involving a simple paramagnetic phase, as discussed below.

The d$M$/d$H$ data for $H$\,$\perp$\,$c$ have maxima at $\mu_0H~\approx$\,0.4\,T and 4.6\,T for $T$\,=\,0.4\,K.  These values coincide well with the $M$($T$) data, thus revealing that these critical fields are associated with the transitions from AFM$_0$ to AFM$_C$ near 0.4\,T and from AFM$_C$ to a partially-polarized paramagnet (or short-range order) at 4.6\,T.  Both of these critical fields decrease slightly upon increasing $T$ toward $T_\mathrm{N}$, as shown in Fig.\,\ref{MH}(c).  At $T$\,=\,0.55 and 0.60\,K, broad maxima are observed in d$M$/d$H$ at $\approx$\,0.75 and 1.66\,T, respectively.

\begin{figure}[b!]
\includegraphics[width=0.92\columnwidth]{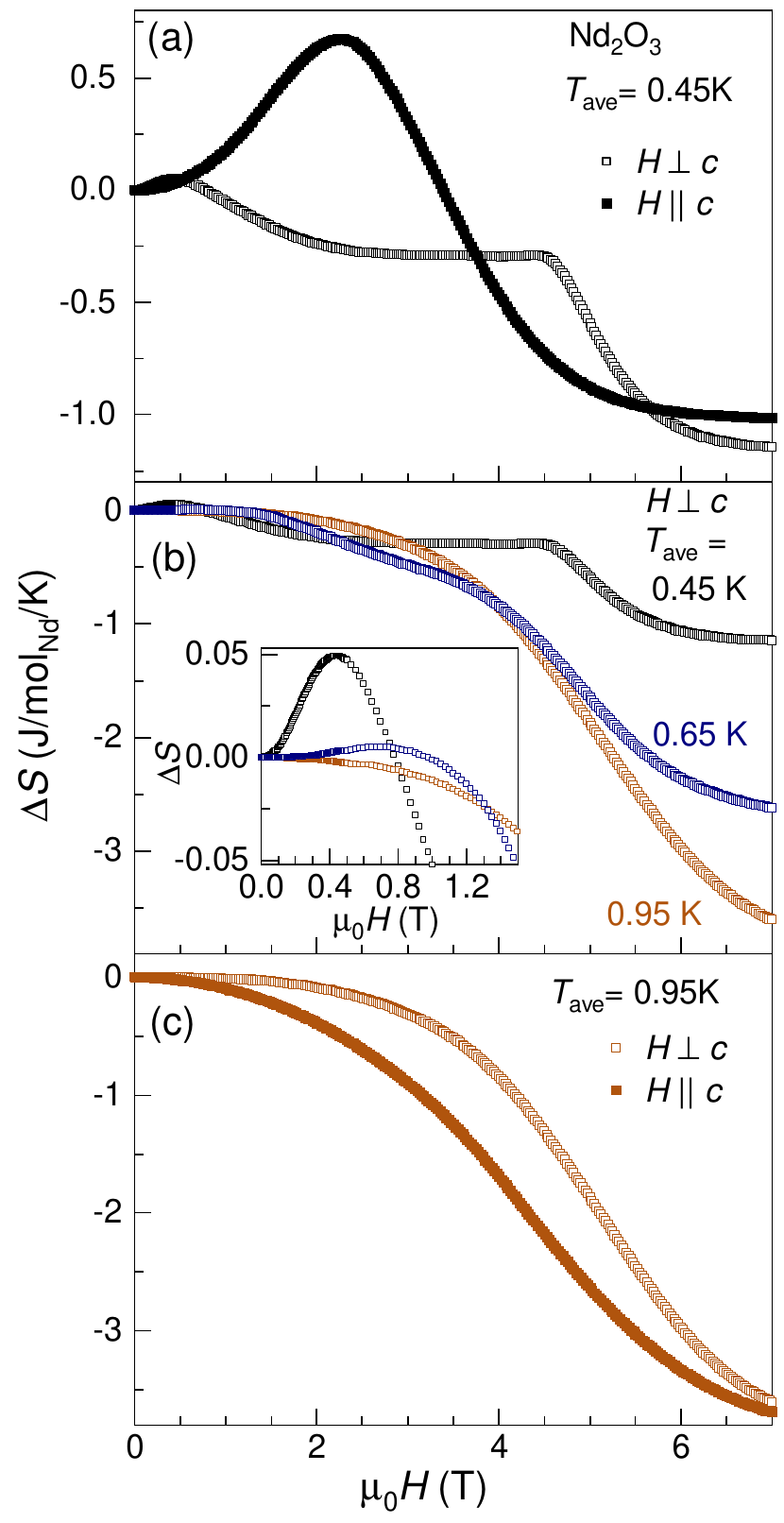}
\caption{\label{Entropy}  Magnetocaloric entropy change as a function of applied field at (a) $T_{ave}$\,= 0.45 K for $H$\,$\perp$\,$c$ and $H$\,$\parallel$\,$c$, (b) $T_{ave}$ = 0.45, 0.65, and 0.95 K for $H$\,$\perp$\,$c$, and (c) $T_{ave}$\,= 0.95 K for $H$\,$\perp$\,$c$ and $H$\,$\parallel$\,$c$. The inset in (b) has the same vertical axis units as the main panel.}
\end{figure}

To further investigate the characteristics of the magnetic phases, the isothermal magnetization was utilized to calculate the magnetic entropy change as a function of applied field.  This was accomplished using the Maxwell relation  $\Delta S$ ($T$, $H$)\,=\,$\int_{0}^{H} ({dM}/{dT})_H$ d$H$  and is sometimes referred to as the magnetocaloric effect/entropy.  Here, the isothermal magnetization data at $T_1$ and $T_2$\,$>$\,$T_1$ were utilized to obtain the integrand via $(M(T_1)-M(T_2))/(T_1-T_2)$.  Integration with respect to the applied magnetic field $H$ produces $\Delta S$ as function of applied field at an average temperature $T_{ave}$\,=\,($T_1$+$T_2$)/2. We emphasize that $\Delta S$ in Fig.\,\ref{Entropy} is an integrated quantity that accounts for all changes in entropy up to a given $H$.  The sign of the magnetic entropy change depends on the nature of the magnetic phase\cite{Tegus2002,Garst2005,Ranke2009,Paige2015}. A positive $\Delta S$ indicates an antiferromagnetic phase, where the application of a magnetic field introduces disorder in the spin configuration that results in an entropy gain.  This is sometimes called an inverse magnetocaloric effect. On the other hand, the application of magnetic field near a ferromagnetic phase transition reduces the entropy by polarizing the moments, resulting in a negative magnetic entropy change as the field is increased.  $\Delta S$ is also negative in the paramagnetic state.  At a given $T$, a peak in $\Delta S$ is thus expected near a phase transition from an AFM state to a FM or PM one.  Interestingly, increases in entropy near field-induced transitions involving chiral spin structures, such as soliton or skyrmion phases, have also been observed but are not particularly relevant here.\cite{Clements2017,Bocarsly2018}

The magnetocaloric $\Delta S$ obtained using $M$($H$) data below and above $T_{\rm N}$ are shown in Fig.\,\ref{Entropy}(a). For $H$\,$\parallel$\,$c$ at $T_{ave}$\,= 0.45 K, a broad and positive maximum of $\Delta S$ is centered around 2.3\,T.  This coincides with the phase boundary between the AFM$_0$ and PM phases. Above $\approx$\,2.3\,T for $H$\,$\parallel$\,$c$, the integrand ${dM}/{dT}$ is negative and eventually the net $\Delta S$ becomes negative and essentially plateaus as the temperature dependence of the polarized moment is negligible at high fields.  For $H$\,$\perp$\,$c$ at $T_{ave}$\,=\,0.45\,K, a positive maximum magnetic entropy change is observed around $\mu_0H\,\approx$\,0.4\,T that coincides with the phase boundary between AFM$_0$ and AFM$_C$ phases (see inset in Fig. \ref{Entropy}(b)). The positive $\Delta S$ values below 2.3\,T for $H$\,$\parallel$\,$c$ and 0.4\,T for $H$\,$\perp$\,$c$ suggests that compensated AFM order exists below these critical fields.  Similarly, the region of negative $\Delta S$ occurring in the field range 0.8\,T - 4.5\,T at $T_{ave}$\,=\,0.45\,K with $H$\,$\perp$\,$c$ reveals the non-compensated nature of the spin configuration where the application of a field is mostly polarizing moments as opposed to causing fluctuations and disorder. Fig. \ref{Entropy}(b) shows the magnetic entropy change for $T_{ave}$\,=\,0.45\,K, \,0.65\,K, and \,0.95\,K for $H$\,$\perp$\,$c$. The positive $\Delta S$ observed at $T_{ave}$\,=\,0.45\,K is suppressed above $T_{\rm N}$ as shown in inset of Fig. \ref{Entropy}(b). The existence of positive $\Delta S$ at $T_{ave}$\,=\,0.65\,K (above $T_{\rm N}$ = 0.55 K) evidences the AFM correlations in these data below about 1\,T. As shown in Fig. \ref{Entropy}(c), the $\Delta S$ at $T_{ave}$\,= 0.95 K is negative and increases smoothly with increasing $H$, and this behavior is consistent with a paramagnetic (or ferromagnetic) phase.

\begin{figure}[h!]
	\includegraphics[clip,width=\columnwidth]{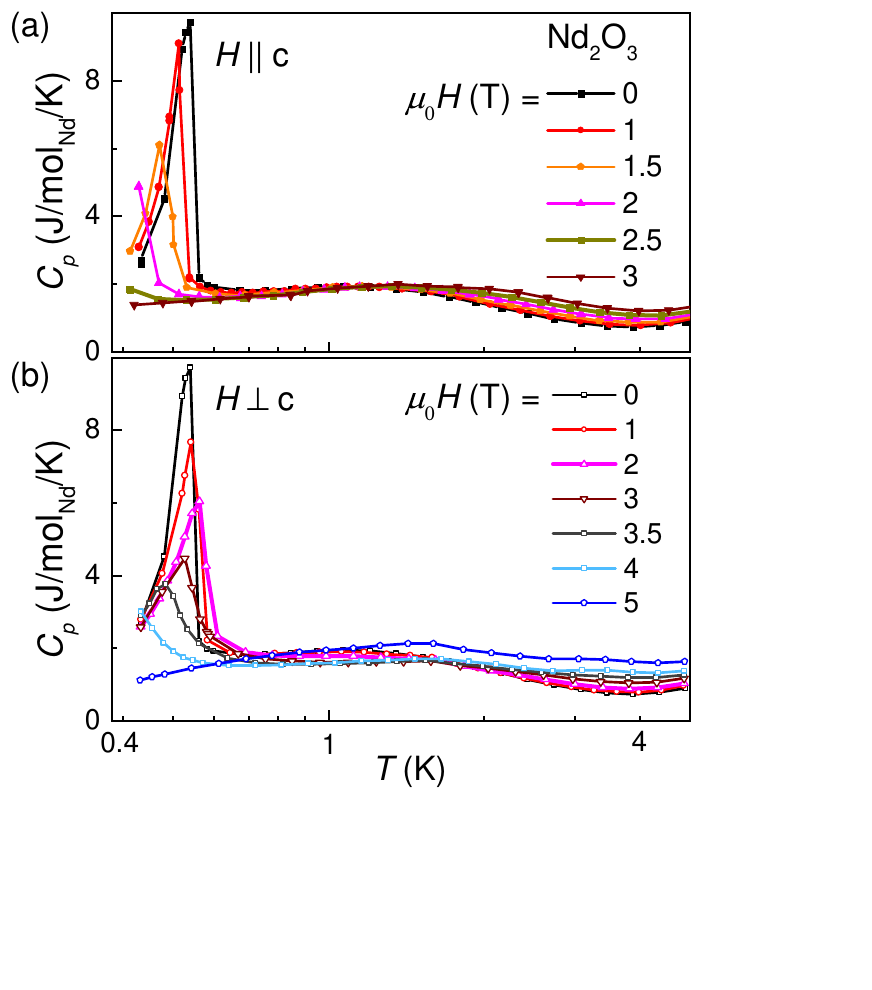}
\caption{\label{Cp} Temperature-dependent specific heat of Nd$_2$O$_3$ with magnetic fields applied (a) along the $c$ axis and (b) within the $ab$ plane.  The horizontal axes are plotted on a log scale.}
\end{figure}

Specific heat measurements were performed down to 0.4\,K to assess the impact of an applied field on $T_{\rm N}$.  The data are shown in Fig. \ref{Cp}, where a sharp $\lambda$-like anomaly is observed at $T_\mathrm{N}$\,=\,0.55\,K for zero-field.  These data have a broad maximum centered around 1.2\,K, which is seemingly caused by static and/or dynamic short-range correlations and is discussed below.  The zero-field behavior observed in $C_P$ for the single crystal is in agreement with our previous results on polycrystalline samples\cite{Sala2018}.   Specific heat data at higher temperatures can be found in the literature.\cite{Goldstein1959,Justice1963} We note that the Schottky anomaly associated with the first excited crystal field level has a broad maximum near 12\,K, as discussed below.

\begin{figure}[t!]
	\includegraphics[clip,width=\columnwidth]{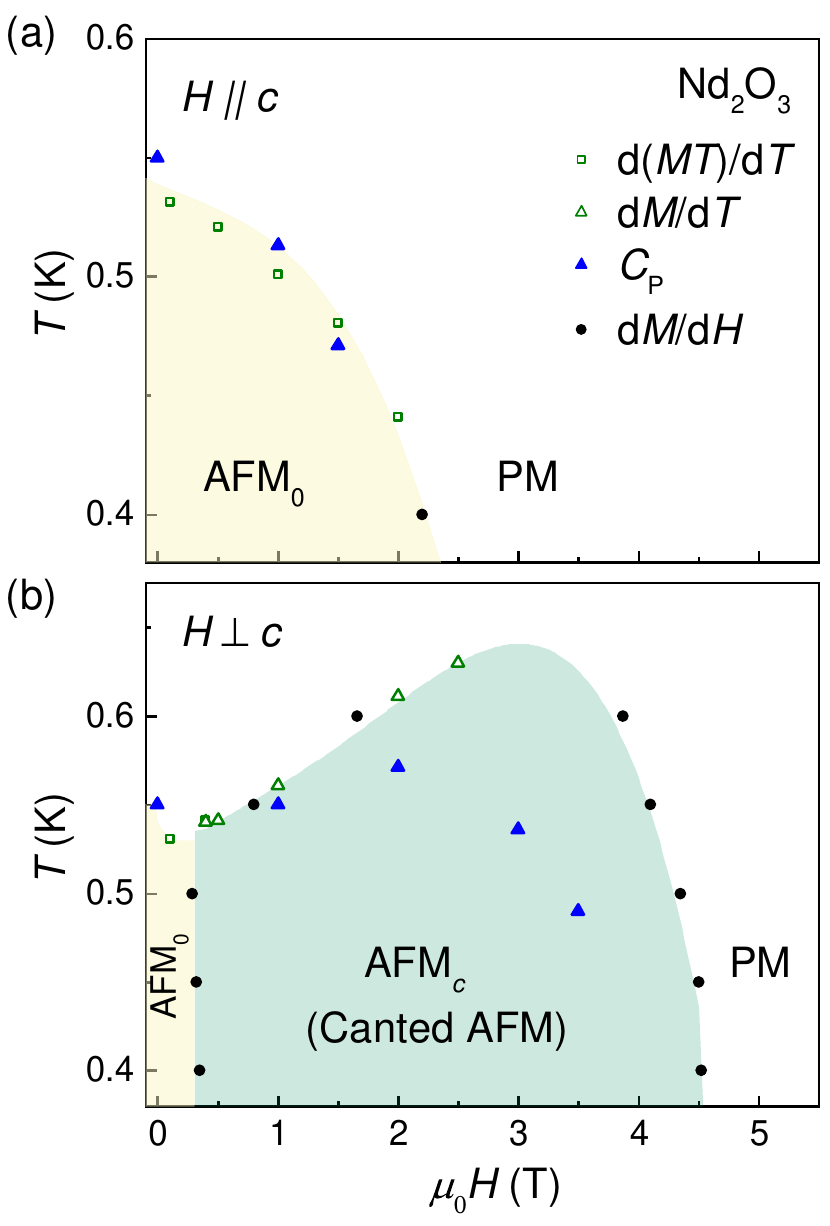}
\caption{\label{Phase} Temperature $T~vs.~$ field $H$ magnetic phase diagram of Nd$_2$O$_3$ for (a) $H//c$ and (b) $H\perp c$ with critical points obtained by various methods as identified in the legend.}
\end{figure}

As shown in Fig.\,\ref{Cp}(a), a magnetic field applied along [001] continually shifts the $\lambda$ anomaly at $T_{\rm N}$ to lower $T$, and the feature is suppressed beyond 0.4\,K around 2.5\,T.  These results are consistent with the magnetization data that revealed a smooth decrease in $T_{\rm N}$ for fields applied along the $c$-axis.  The specific heat data for $H$\,$\perp$\,$c$ are shown in Fig.\,\ref{Cp}(b), and it is observed that the peak of the $\lambda$ anomaly is slightly higher in temperature for $\mu_0H$\,=\,1 and 2\,T than for zero-field.  This is qualitatively consistent with the magnetization data, though the shift in the inflection of $M$($T$) is more significant than that in the anomaly of the specific heat.  The $\lambda$ anomaly is suppressed beyond 0.4\,K by 4-5\,T for $H$\,$\perp$\,$c$. 

The temperature and field-induced phase transitions observed via magnetization and specific heat measurements are summarized in the $H$-$T$ diagram shown in Fig.\,\ref{Phase}.  This phase diagram does not include the short-range order inferred from these thermodynamic data and significant spin-spin correlations are expected in the paramagnetic (PM) phase.  The phase AFM$_0$ is a collinear antiferromagnetic ground state with moments in the $ab$-plane.\cite{Sala2018}  $T_{\rm N}$ for this phase is continually suppressed with increasing field for $H$\,$\parallel$\,$c$.  The phase AFM$_C$ exists only for $H$\,$\perp$\,$c$ and is likely a canted phase based on the behavior of $M$($T$). The shaded areas are based on features in the magnetization data, and a smaller area is implied by the locations of $\lambda$ anomalies in the specific heat data (for $H$\,$\perp$\,$c$).  In the AFM$_0$ region, $T_{\rm N}$ was defined as the peak in d($MT$)/dT, while in AFM$_C$ the peak in d$M$/d$T$ was utilized due to the ferromagnetic-like nature of $M$($T$)\cite{Fisher1962}.  Maxima in d$M$/d$H$ were also used to identify critical temperatures, though the pertinent features are broad for certain $H$,$T$ and a simple maximum could not always be selected (see Fig.\,\ref{MH}(c)).  It is challenging and likely nonphysical to demarcate regions associated with the short-range order, and thus short-range order is not indicated on the phase diagram.

\subsection{Short-Range Correlations Above $T_{\rm N}$}

The specific heat is characterized by a broad feature centered above 2$T_{\rm N}$.  This was previously shown to contain around 70\% of the expected entropy ($R$ln2) for the magnetic doublet ground state.\cite{Sala2018}  The extent of this entropy release is taken as evidence for the existence of strong spin-spin correlations above $T_{\rm N}$.

The field-dependent specific heat data in Fig.\,\ref{Cp} were analyzed to obtain the magnetic entropy as a function of field and temperature.  This temperature-dependent view of entropy is different from the magnetocaloric entropy that was utilized to characterize the field-induced entropy at a given temperature for assessing the nature of the spin-spin alignments.  The magnetic specific heat was obtained via $C_{mag}$\,=\,$C_{Nd_2O_3}-C_{La_2O_3}$ where $C_{La_2O_3}$ is the specific heat of non-magnetic, isostrucural La$_2$O$_3$ that was taken as a phonon background.  The magnetic entropy was then obtained as $S_{\textrm{mag}}$($T$)\,=\,$\int_{0}^{T} (C_{mag}/T) dT$ from $T$ = 0.4 to 5\,K.

\begin{figure}[h!]
	\includegraphics[clip,width=\columnwidth]{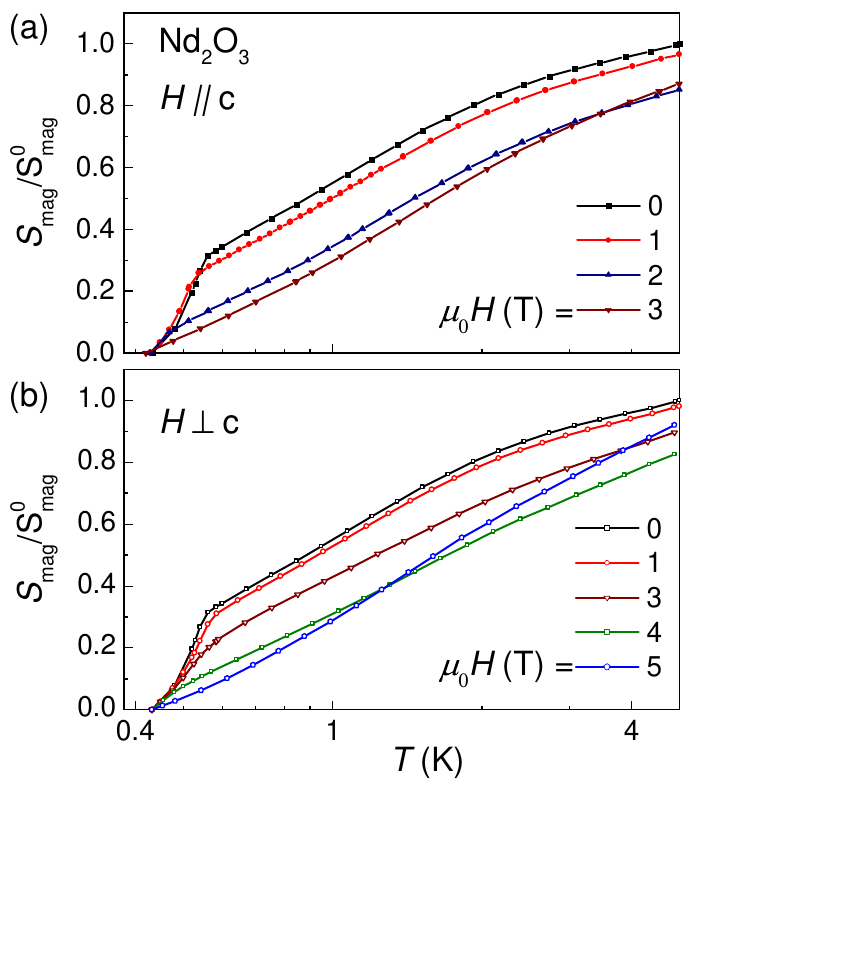}
\caption{\label{CpEntropy} Magnetic entropy as a function of temperature for various applied field, with data normalized to the maximum value obtained in zero-field.  The applied field is (a) along the $c$ axis and (b) within the $ab$ plane.}
\end{figure}

The calculated magnetic entropy is shown in Fig.\,\ref{CpEntropy}. As expected based on previous work, the majority of the entropy is not released until well-above $T_{\rm N}$.  Obtaining the complete $R$ln(2) entropy would seemingly require integration above 5\,K where the higher-temperature Schottky anomaly starts to hinder analysis; data below 0.4\,K will also contribute some missing entropy.  As such, the effect of applied field is examined by normalizing the integrated entropy at various fields to the value $S_{mag}^{0}$ obtained by integrating the zero-field data up to 5\,K.  Increasing the field, regardless of orientation, decreases the entropy released at low $T$ by suppressing the anomaly associated with long-range AFM$_0$ order.

The position of the broad maximum in $C_P$ associated with short-range correlations shifts to slightly higher $T$ with applied fields, as shown in Fig.\,\ref{Cp}.  The role of crystal field level splitting due to the applied field is unclear, but likely does not significantly impact the data below 4-5\,K.  The shape of the broad maximum is similar to that of a two-level Schottky system, though the magnitude of the maximum in the experimental $C_{mag}$ is smaller than that expected from a simple Schottky model.  Based on the inelastic neutron scattering data discussed below, we believe that this broad maximum in the specific heat is not associated with a simple two-level system but is instead associated with dynamic spin-spin correlations that persist to temperatures well-above $T_{\rm N}$.

Neutron scattering experiments were undertaken in an effort to understand if the entropy-released well-above $T_{\rm N}$ is associated with static or dynamic spin-spin correlations.  As mentioned above, a single crystal was examined on the CORELLI beamline at the SNS.  This experiment was performed in an attempt to observe diffuse scattering at the position of Bragg reflections associated with the long-range magnetic order that possesses a propagation vector $\vec{k}\,=\,(\frac{1}{2},0,\frac{1}{2})$.  Inspection of the data above $T_{\rm N}$ did not reveal diffuse scattering at associated reciprocal space locations and thus the data do not provide evidence for static short-range order. However, diffuse magnetic scattering can be challenging to detect.  In addition, the data do not preclude short-range magnetic order with scattering at $k$\,=\,0 locations where strong intensity from the crystal lattice dominates the signal.

\begin{figure}[b!]
	\includegraphics[width=0.9\columnwidth]{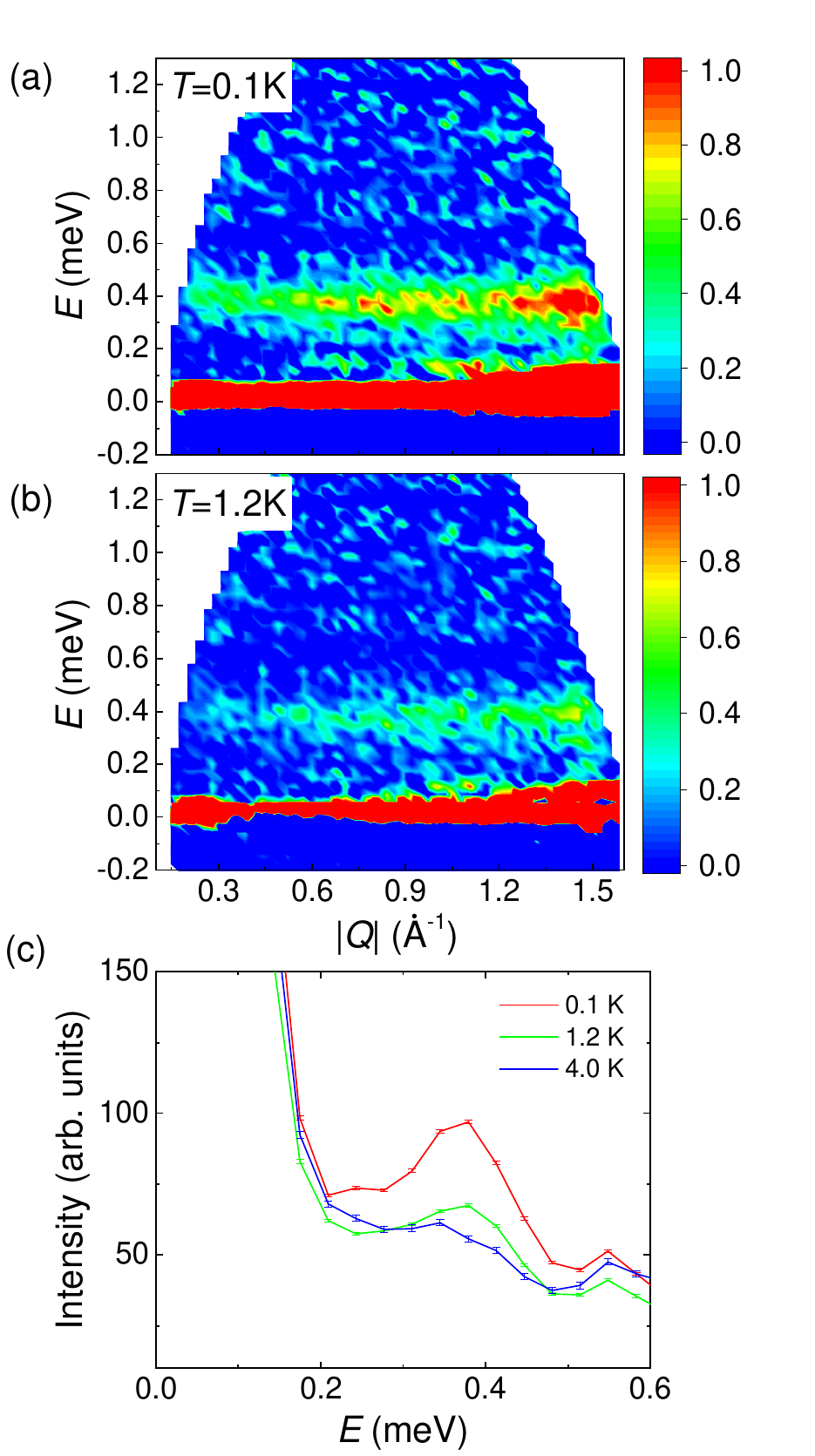}
\caption{\label{INS1} Low-energy spectra for Nd$_2$O$_3$ obtained using inelastic neutron scattering on single crystals, with data collected at (a) $T$\,=\,0.1\,K and (b) 1.2\,K. For both (a) and (b) the data at 4\,K were used as the background and subtracted from the data at each temperature.  The data are presented as a pseudo-powder average of the momentum transfers $Q$ that were probed.  (c) Energy-dependence of the intensity, integrated across 0.2\,$\leq$\,$|Q|$\,$\leq$\,1.5\AA$^{-1}$ at the three temperatures examined.}
\end{figure}

Inelastic neutron scattering experiments were performed on the Cold Neutron Chopper Spectrometer (CNCS beamline at SNS).    The inelastic spectra below and above $T_{\rm N}$ are shown in Fig.\,\ref{INS1}(a,b), where the data are presented in color maps as a function of energy transfer $E$ and the magnitude of the momentum transfer $Q$. The data in Fig.\,\ref{INS1}(c) were obtained by integrating over 0.2\,$\leq$\,$|Q|$\,$\leq$\,1.5\AA$^{-1}$.  The incident energy was 1.72\,meV with a corresponding energy resolution of 40 $\mu$eV at the elastic line. To improve counting statistics, the single crystal data were binned according to $E$ and $|Q|$.  The resulting spectral maps are similar to a pseudo-powder average due to the large area of reciprocal space covered during the measurement. 

The excitation spectrum in Fig.\,\ref{INS1}(a) has a mode(s) with a characteristic energy of $\approx$\,0.37\,meV at 0.1\,K (below $T_{\rm N}$).    Interestingly, the scattering persists upon warming through $T_{\rm N}$ to 1.2\,K with a small amount of intensity remaining at $T$\,=\,4\,K.  The energy of maximum intensity is not strongly dependent upon temperature, as shown in Fig.\,\ref{INS1}(c).

The $Q$-dependence of the intensity, the evolution with temperature, and the characteristic energy together suggest that this scattering is due to collective spin-spin correlations of magnetic origin.  The intensity increases with increasing $Q$ for the reciprocal space covered in this experiment.  While the intensity of lattice excitation (phonon) would also increase with increasing $Q$, a phonon at this energy would be expected to increase in intensity on warming due to thermal population in the range of temperatures presented.  The form factor for magnetic scattering actually causes the intensity to decrease with increasing $Q$.  However, in addition to the form factor, spin-spin correlations introduce additional modulations to the scattering intensity.  A well-known example is the scattering intensity of spin waves from an antiferromagnetic spin configuration where the maximum intensity does not occur at $Q$=0 as would be expected from the form factor alone.  The fact that the intensity does not follow the form factor indicates that this scattering is associated with collective behavior and not a single ion effect such as a crystal field excitation.  From this perspective, we expect that the intensity in  Fig.\,\ref{INS1}(a,b) would decrease after peaking at a particular $Q \geq$ 1.45 \AA\, if sufficient data were available.  As such, these data seem to indicate that Nd$_2$O$_3$ hosts dynamic spin-spin correlations that persist to well-above $T_{\rm N}$.

The $Q$-integrated data in Fig.\,\ref{INS1}(c) reveal that the scattering has a width of approximately 0.1\,meV and that the position does not change significantly between 0.1 and 1.2\,K.  This low-energy scattering is most intense below $T_{\rm N}$, and thus it appears to be a characteristic feature of the spin wave spectrum in the ordered state.  Given that static short-range order has not been evidenced (via diffuse scattering), the physical origin of the persistence of the low-energy scattering above $T_{\rm N}$ is unclear.  It is important to reiterate that the inability to detect short-range order through diffraction experiments does not mean that (static) short-range correlations do not exist.  Still, this is a surprising result and suggests that the frustrated nature of the Nd$_2$O$_3$ lattice leads to strong fluctuations. These dynamic correlations impact the thermodynamic properties, most notably the specific heat.  The intensity of this scattering could be tracked as a function of temperature to provide an order parameter and this may allow for more detailed modeling of the specific heat data. Such an investigation would allow the phase diagram in Fig.\,\ref{Phase} to be modified and could reveal additional phase boundaries depending on the type of magnetic frustration.\cite{Hoang2012,Liu2016}  The behavior of this scattering in an applied field may be especially informative.

The lack of dispersion suggests that the states are relatively localized, though the finite peak width being larger than the instrumental energy resolution implies the existence of a potential dispersion.  Similar modes have been observed in cluster-based systems, such as the breathing pyrochlore Ba$_3$Yb$_2$Zn$_5$O$_{11}$ with non-interacting Yb tetrahedra.\cite{Rau2016,Haku2016}  However, the three-dimensional nature of Nd$_2$O$_3$ makes it difficult to identify a structural feature that would produce a localized excitation similar to that in cluster-based systems.  Flat modes are often a sign of Ising interactions\cite{Ikeda1978}.  The data were, unfortunately, found to be unsuitable for modeling due to poor statistics and a large number of symmetry-allowed terms in the Hamiltonian.  Similarly, a larger $Q$-range would allow greater constraints to be placed on any models considered.  Therefore, at this time, the exact nature of the low-energy mode remains unclear.

\section{Summary and Outlook}

Nd$_2$O$_3$ is an easy-plane antiferromagnet with long-range order below $T_{\rm N}$\,=\,0.55\,K. The material has a three-dimensional crystal structure that promotes frustrated and/or competing interactions due to short distances between stacked triangular arrays. The magnetization near $T_{\rm N}$ strongly depends on the orientation and magnitude of the applied magnetic field. Fields applied along the $c$ axis (perpendicular to the spins) gradually suppress $T_{\rm N}$ to below 0.4\,K, with a critical field of roughly 2.5\,T.  When the field is applied within the basal plane, a qualitative change in $M$($T$) is observed for fields above 0.4\,T. The data suggest the development of a canted antiferromagnetic structure with a small, uncompensated moment.  The exact nature of this phase needs to be addressed through neutron diffraction experiments in an applied field.  The phase transitions observed by thermodynamic measurements have been summarized in field-temperature phase diagrams. A limitation of the phase diagrams exists due to the presence of short-range correlations.

The existence of short-range correlations above $T_{\rm N}$ was previously speculated based on specific heat measurements on polycrystalline Nd$_2$O$_3$ that revealed a large portion of the ground state magnetic entropy is released well-above $T_{\rm N}$. The same behavior has now been observed in single crystals of Nd$_2$O$_3$, suggesting this is an intrinsic behavior associated with the frustrated lattice geometry.  This work explored the possible existence of short-range correlations via neutron scattering.  Elastic neutron diffraction experiments were not able to evidence static short-range order above $T_{\rm N}$. Instead, inelastic measurements proved essential and a low-energy, flat mode(s) was observed to persist at 1.2\,K.  The scattering is centered near 0.37\,meV at 0.1\,K and this energy does not change much with $T$.   The connection between the persistent, low-energy scattering and the underlying Hamiltonian needs to be addressed to determine if Nd$_2$O$_3$ is a model material to study the physics of a (frustrated) stacked triangular lattice. Further inelastic scattering experiments are required to obtain data sufficient to address the nature of the magnetic Hamiltonian.

\section{Acknowledgments}

We thank G. Halasz and H. Zhang for useful comments and discussions. This research was supported by the U.S. Department of Energy, Office of Science, Basic Energy Sciences , Materials Science and Engineering Division. Work at the Oak Ridge National Laboratory Spallation Neutron Source was supported by U.S. DOE, Office of Science, BES, Scientific User Facilities Division.


%

\end{document}